\documentstyle[twoside,fleqn,espcrc2,epsf,amssymb]{article}

\newcommand{\noi}{\noindent}
\newcommand{\ra}{\rightarrow}
\newcommand{\vs}{\vspace}
\newcommand{\eq}{\begin{equation}}
\newcommand{\en}{\end{equation}}
\newcommand{\eqa}{\begin{eqnarray}}
\newcommand{\ena}{\end{eqnarray}}

\newcommand{\bpsi}{{\bar \psi}}

\newcommand{\AmS}{{\protect\the\textfont2
  A\kern-.1667em\lower.5ex\hbox{M}\kern-.125emS}}

\title{
{
\vspace{-2.0cm} 
\scriptsize
\hfill
\parbox{30mm}{
DESY 98-161\\HLRZ1998\_61\\BI-TP 98/32\\HU-EP-98/68
             }
}\\[5mm]
Passing through the `chiral limit' in quenched QCD with 
Wilson fermions}

\author{A. Hoferichter\address{DESY/NIC(HLRZ) Zeuthen, Germany} 
        \thanks{Talk given by A. Hoferichter.},
        E. Laermann\address{Fakult\"at f\"ur Physik,
				     Universit\"at Bielefeld, Germany},
	  V.K. Mitrjushkin\address{Joint Institute for Nuclear Research, 
					Dubna, Russia},
        M. M\"uller-Preussker\address{Institut f\"ur Physik, 
						Humboldt-Universit\"at
						zu Berlin, Germany}
	and P. Schmidt$\mbox{}^{\mbox{\scriptsize b}}$
}
       
\begin{document}

\begin{abstract}
We investigate the limit of vanishing quark mass in quenched lattice QCD
with unimproved Wilson fermions at $\,\beta=6.0$.
Exploiting the correlations of propagators at
different time slices we extract pion masses extremely
close to the `chiral limit', despite the presence of `exceptional 
configurations'.
With this at hand, the existence of quenched chiral logarithms can be
demonstrated, provided, finite size effects are small.
With reference to the phase diagram proposed by Aoki \cite{aoki} also 
the range  $\,\kappa > \kappa_c\,$ is investigated.
The width of a potential parity-flavor violating phase can, if
at all, hardly be resolved.
\end{abstract}

\maketitle

\section{INTRODUCTION}

A recent investigation \cite{sharpe0} pointed out that two basic
scenarios are possible in the case of two-flavor QCD with Wilson 
fermions when the quark mass is tuned to zero in the 
thermodynamic limit:

\begin{itemize}
\item[(a)] Aoki phase ($\,\mathbb{A}\,$)

 -- there exists a phase in which parity-flavor symmetry gets
    spontaneously broken at {\sl finite} lattice spacing $\,a$ --
    this phase is characterized by two corresponding massless 
    Goldstone bosons $\,\pi^{\pm}\,$, whereas $\,\pi^0\,$ is  
    massive, except on the phase boundaries,

 -- in the limit $\,a \ra 0\,$ the pattern of parity-flavor 
    symmetry breaking can be identified with the `genuine' chiral 
    symmetry breaking, 
          
 -- for small $\,a\,$, the width of $\,\mathbb{A}\,$ should scale 
    as $\,\Delta \frac{1}{\kappa}  \sim \, a^3\,$ (up to log. corrections)
          
\item[(b)] 

there is no symmetry breaking involved in the `chiral limit'
   at {\sl finite} lattice spacing,

-- correspondingly, this case would be indicated by equal,
   non vanishing pion masses in the limit $\,\kappa \ra \kappa_c$; (but under
   certain conditions one can expect $\,m_{\pi} \sim O(a)$) 
\end{itemize}

\noi
It is a priori not known, whether both scenarios can coexist in the 
phase diagram (as a function of $\,\beta$). 
Based on quenched numerical
observation \cite{aoki3}, the Aoki scenario is reported to hold at least
up to $\,\beta \simeq 5.7$.
In the full case \cite{bitar}, the 
existence of $\,\mathbb{A}\,$ was shown up to $\,\beta \sim 4$, the reported
absence of $\,\mathbb{A} \,$ for $\,\beta \geq 5\,$ could also be 
interpreted as a sign for a narrow Aoki phase at larger $\,\beta$. 
For the finite temperature case, see \cite{finiteT}.

In fact, all investigations of the low-lying spectrum are based on 
the {\sl assumption}, that, at least in the thermodynamic limit, the 
pion mass(es) will eventually vanish at some value of $\,\kappa$ in accordance 
with PCAC. This is surely true in the continuum limit. But in general,
a behavior like $\,m_{\pi}^2 \, \propto \, m_q\,$ for small $\,m_q\,$ does not
automatically guarantee the existence of a {\sl chiral limit} in a
theory with matter fields realized by Wilson fermions, since other symmetries
may take over the r\^ole of chiral symmetry at finite $\,a$.
In this respect it is crucial to obtain direct information from the vicinity of
$\,\kappa_c(\beta)$, as it is the aim of this study. In addition, we address the
question of quenched chiral logs.

The model under consideration is 
standard Wilson lattice QCD with action
$S = S_G + S_F$, $S_G$ denoting
the plaquette $SU(3)$ gauge part and $S_F$   
the fermionic action\,:
\eqa
S_{F} =  \sum_{x,y}
\bpsi_x {\mathbb{M}}_{xy} \psi_y\,,
\quad 
{\mathbb{M}} = 
\hat 1 \, - \, \kappa \, {\mathbb{D}} \,, \nonumber \\
{\mathbb{D}}_{xy} =  \sum_{\mu} \left [ \delta_{y, x+\hat{\mu}} 
P^{-}_{\mu} U_{x \mu}
+ \delta_{y, x-\hat{\mu}} P^{+}_{\mu} U_{x-\hat{\mu},\mu}^{\dagger} 
\right ]
\ena
where $P^{\pm}_{\mu}={\hat 1}\pm\gamma_{\mu}$.
On each configuration, we use the following 
(non-singlet) correlator in order to extract the 
$\,\pi\,$ mass \,: 

\eq
\Gamma_{\pi}(\tau) \, \sim \,
\sum_{\vec{x},\vec{y}}
\mbox{Tr}
\left( {\cal M}^{-1}_{xy} \gamma_5 {\cal M}^{-1}_{yx}
\gamma_5 \right)
\en
\noi with $\tau = x_4 - y_4$.
The current data sample consists of $\,O(100)\,$ quenched measurements 
each on $\,16^3\times32\,$ and $\,8^3\times32\,$ lattices
at $\,\beta=6.0$. As inversion methods, BiCGstabI (CG)
have been used for $\,\kappa < \kappa_c\,$ ($\,\kappa > \kappa_c$).

%
%
%
\begin{figure}[h]
\vspace*{-1.75cm}
\begin{center}
\hbox{
\epsfysize=8.7cm
\epsfxsize=7.5cm\epsfbox{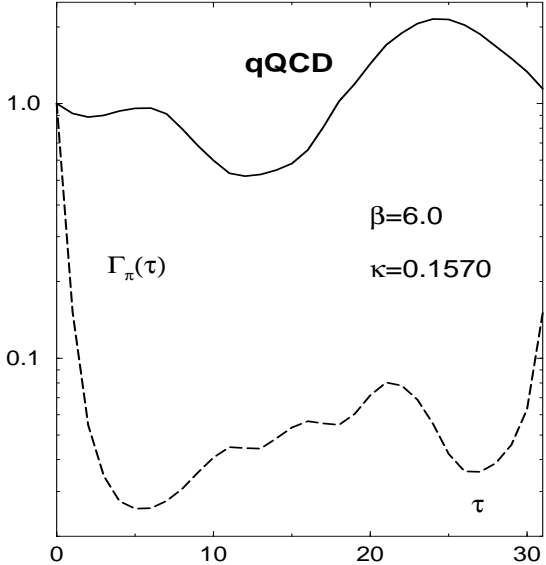}
     }
\end{center}
\vspace*{-1.8cm}
\caption{Typical `exceptional' $\,\pi\,$ propagators which
have been part of the data sample.}
\label{fig:Pi_Prop}
\end{figure}
\vs*{-1.0cm}
\section{THE VICINITY of $\,\kappa_c(\beta=6.0)$}
\label{sect:vicinity}

The quenched approximation is featured
by `exceptional configurations' appearing at small 
values of the quark mass.
Typical examples of `exceptional' $\,\pi\,$ propagators which
have been
%
%
%
%
%
\protect\begin{figure}[t]
\vspace{-0.85cm}

\begin{center}
\hbox{
\epsfysize=8.5cm
\epsfxsize=7.5cm\epsfbox{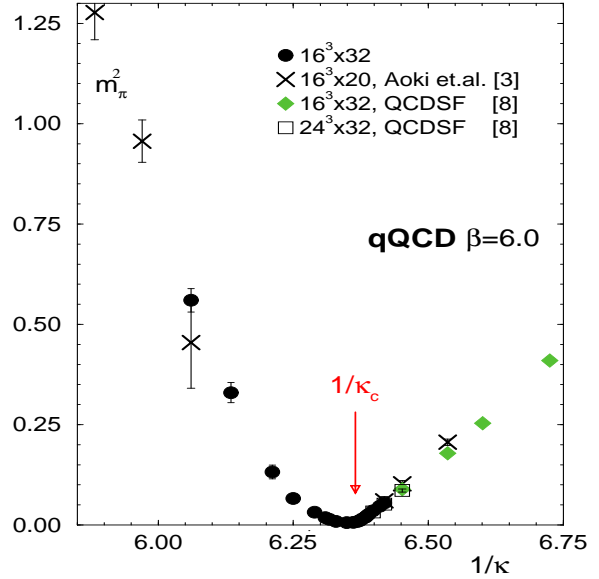}
     }
\end{center}
\vspace*{-1.8cm}
\caption{View of $\,m^2_{\pi}\,$ vs. $1/\kappa$ at 
$\,\beta=6.0$. 
}
\label{fig:Global_mPi}
\vs*{-0.9cm}
\end{figure}
%
%
%
part of the data sample are given in Fig.\ref{fig:Pi_Prop}.
It is common practice to discard them `by hand' from the 
sample.
But, viewed in a sufficiently large ensemble the individual 
$\,\Gamma_{\pi}$'s (including `exceptional' cases) obey a
special form of linear correlation at different time slices.
In this case we apply a variance reduction technique
\cite{hmm95} taking advantage of
$\,\left < x/y \right > = \left < x \right > / \left < y \right >$, with
$x = \Gamma_{\pi}(\tau+1)$, $y=\Gamma_{\pi}(\tau)$.
Without reducing the data sample this allows  
to extract physical information in a region of extremely
small $\,m_q$.
For other approaches to the problem of `exceptional
configurations', see \cite{eichten}.

In Fig.\ref{fig:Global_mPi} we display the behavior of
$\,m_{\pi}^2\,$ vs. $\,1/\kappa\,$ at $\beta=6.0$.
This quantity represents the (mass)$^2$ of the $\,\pi^{\pm}$, which, in full 
($N_f=2$) QCD, should vanish linearly in $\,1/\kappa\,$
for $\,\kappa \ra \kappa_c\,$ and remain zero within 
$\,\mathbb{A}$. This behavior is 
predicted \cite{sharpe0} to be symmetric with respect to 
the approach to the phase boundaries of $\,\mathbb{A}\,$
from outside. 
This is not the case as seen in Fig.\ref{fig:Global_mPi}.
Since the scenario depends on $\,N_f$, quenching effects 
might be large - there is need to quantify them here.
At present, there is no theoretical handle to extrapolate 
$\,m^2_{\pi} \ra 0\,$ for the data at $\,\kappa > \kappa_c$.
Even if one would enforce a linear dependence as has been done in \cite{aoki3}
the modulus of the slope would be different from that found for $\,\kappa < \kappa_c$.
Hence, the width of a potential Aoki phase cannot be firmly resolved at 
$\,\beta=6.0$. This, however, is not in contradiction with 
$\,\Delta \frac{1}{\kappa}  \,\sim \, a^3$. 
For the minima of $\,m_{\pi}^2 \,$ the existence of $\,\mathbb{A}\,$ 
would imply $\,m_{\pi, min}^2 = 0\,$ for infinite volume.
This is supported by a comparison with 
$\,8^3\times32$ data\footnote{not displayed}, but has to be checked on 
various lattice sizes. 

\section{QUENCHED CHIRAL LOGS}
\label{sect:qXlogs}

From the discussion in sect.\ref{sect:vicinity} it 
is justified to assume the presence of 
quenched {\sl chiral} logarithms in the data 
below $\,\kappa_c$.
Based on quenched chiral perturbation theory one 
expects \cite{Xpt}\,:

\eq
\ln \left ( 
\frac{m^2_{\pi}}{m_q} 
\right ) 
= c_0 - \frac{\delta}{\delta +1} \ln m_q 
 + c_1 m_q + c_2 m^2_q
                                   \label{xpt_mpi}
\en

\begin{figure}[th]
\vspace{-0.85cm}
\begin{center}
\hbox{
\epsfysize=8.7cm
\epsfxsize=7.5cm\epsfbox{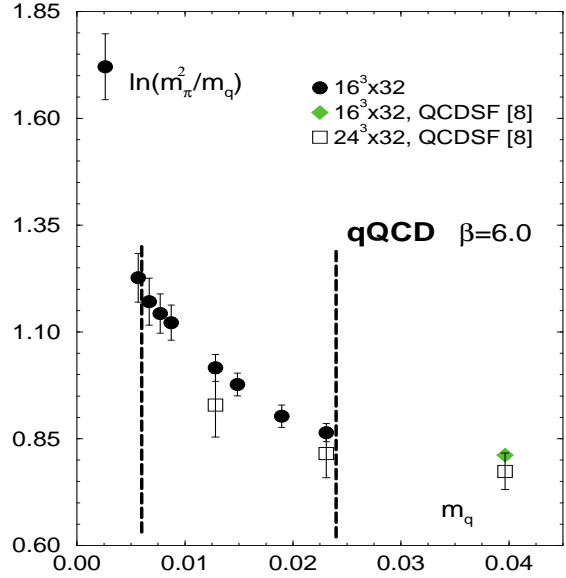}
     }
\end{center}
\vspace*{-1.8cm}
\caption{
$\ln(m_{\pi}^2/m_q)\,$ as a function of $\,m_q$.
}
\label{fig:delta}
\end{figure}

\noi with $\,\delta = m_0^2/(24\pi^2\,f_{\pi}^2)\, > 0$. 
Fig.\ref{fig:delta} represents $\,\ln(m_{\pi}^2/m_q)$ as a function of
$\,m_q$.
For very small $\,m_q\,$ the term $\,\propto\, \ln m_q\,$ should dominate, which 
is seen in our data points for $\,\kappa < \kappa_c$.
The result for $\,\delta\,$ sensitively depends on the determination 
of $\,m_q$. It is more reasonable to eliminate $\,\kappa_c\,$ entering
eq.(\ref{xpt_mpi}) as a free parameter by using directly the PCAC quark 
mass, if possible (cf. \cite{dirk}). 
Preliminary, we fixed $\,\kappa_c\,(=0.15693)\,$ by a 
$\,\chi^2\,$ fit to our 7 data points in the range 
indicated by the the vertical lines in Fig.\ref{fig:delta} 
assuming $\,c_1 = c_2=0$ and obtaining $\,\delta\,\sim \,0.3$ 
as a first estimate.
The data points at larger values of $\,\kappa\,$ have been omitted, because of
expected finite size effects.

We can draw the qualitative conclusion, that our data is compatible 
with quenched chiral logarithms, with a value of $\,\delta\,$ in the 
expected range. 
However, it requires more investigation in order to come to a quantitative
conclusion. 

\vs*{-0.028cm}
\section{SUMMARY}

We conclude, that at the given $\,\beta\,$ value 
a finite width of the Aoki phase cannot be firmly resolved.
At the same time, quenched {\sl chiral} logarithms are visible 
in our data.

This work was supported by the grant  
TMR ERB FMRX-CT97-0122.

%
%
%
%
%
%
%
%
%

%

\end{document}